\documentclass[%
reprint,
amsmath,amssymb,
aps,
prb,
floatfix,
showpacs
]{revtex4-1}

\usepackage{makecell}
\usepackage{graphicx}
\usepackage{dcolumn}
\usepackage{bm}

\newcommand{\Ni}{NiWO$_4$}

\bibpunct{[}{]}{;}{n}{}{}

\begin{document}


\title{Lattice and spin dynamics in a low-symmetry antiferromagnet \Ni{} }

\author{M. A. Prosnikov}
 \email{yotungh@gmail.com}
 \affiliation{Ioffe Institute, Russian Academy of Sciences, 194021 St.-Petersburg, Russia}
\author{V. Yu. Davydov}
 \affiliation{Ioffe Institute, Russian Academy of Sciences, 194021 St.-Petersburg, Russia}
\author{A. N. Smirnov}
 \affiliation{Ioffe Institute, Russian Academy of Sciences, 194021 St.-Petersburg, Russia}
\author{M. P. Volkov}
 \affiliation{Ioffe Institute, Russian Academy of Sciences, 194021 St.-Petersburg, Russia}
\author{R. V. Pisarev}
 \affiliation{Ioffe Institute, Russian Academy of Sciences, 194021 St.-Petersburg, Russia}
\author{P. Becker}
 \affiliation{University of Cologne, Sect. Crystallography, Institute of Geology and Mineralogy, 50674 Cologne, Germany}
\author{L. Bohat\'y}
 \affiliation{University of Cologne, Sect. Crystallography, Institute of Geology and Mineralogy, 50674 Cologne, Germany}

\date{\today}

\begin{abstract}
Lattice and magnetic dynamics of \Ni{} single crystals were studied with the use of polarized Raman spectroscopy in a wide temperature range of 10--300~K including the antiferromagnetic ordering temperature $T_N$=62~K.
Static magnetic measurements were used for characterizing the single crystals.
All Raman-active phonons predicted by the group theory were observed and characterized.
Magnetic symmetry analysis was used to determine possible magnetic space groups for \Ni{} which can be also applied to any other isostructural crystal with the same magnetic propagation vector \textbf{k}=(1/2,0,0).
Though the magnetic structure of \Ni{} is relatively simple, a rich set of narrow and broad magnetic excitations with different polarization properties and temperature behavior in the very broad frequency range of 10--200~cm$^{-1}$ was observed, with some modes surviving at temperatures much higher than $T_N$ up to 220~K.
Part of the magnetic excitations were identified as acoustic and optical spin-wave branches which allow us to construct exchange structure and estimate exchange and anisotropy constants with the use of linear spin-wave theory.
Since the magnetic structure can be described as exchange-coupled AFM chains of \textit{S}=1 ions, previously unobserved magnetic excitation at 24~cm$^{-1}$ is tentatively assigned to a Haldane gap mode.
\end{abstract}

\pacs{78.30.Hv, 75.50.Ee, 75.30.-m, 71.15.Mb}

\maketitle


\section{\label{sec:intro}Introduction}
Complex magnetics with particular types of crystallographic structures, frustrated interactions, nontrivial symmetric, antisymmetric and anisotropic exchange interactions within and between sublattices, multisublattice types of exotic magnetic orderings, low-dimensionality, spin dimerization and other specific features are among the most actively studied subjects in the contemporary magnetism of strongly correlated systems. Some examples may be cited, namely the Kitaev model \cite{thio_antisymmetric_1988,chun_direct_2015,reuther_finite-temperature_2011}, the kagome lattices \cite{greedan_geometrically_2001,lee_emergent_2002}, the frustrated spin-ice pyrochlore magnetics \cite{gardner_magnetic_2010}, etc. Magnetics with the spin \textit{S}=1 and larger can show uncommon behavior because of their exotic, biquadratic, and four-spin types of exchange interactions.
There is also an unique branch of low-dimensional magnetics with integer spin - Haldane magnets \cite{affleck_quantum_1989,buyers_experimental_1986}. 

Each of the mentioned systems may possess various kinds of admissible spin-related excitations, sometimes unique, such as spin-gap dimers, multi-magnon or magnetic bound states \cite{lemmens_magnetic_2003,hagiwara_spin_2005}. Complexity of magnetic excitation spectra requires application for their studies various complementary experimental techniques such as microwave, terahertz and infrared spectroscopy of absorption and reflection, Raman scattering spectroscopy, inelastic neutron scattering, and others. In recent years, the ultrafast pump-probe spectroscopy of magnetic insulators with the use of femtosecond terahertz and optical pulses attracts more and more attention, opening new perspectives in ultrafast control equilibrium and non-equilibrium of magnetic states of the media \cite{kirilyuk_ultrafast_2010,kalashnikova_ultrafast_2015}.


There is an interesting group of magnetic insulators of the \textit{A}WO$_4$ type, where \textit{A} = Mn, Fe, Co, Ni and Cu \cite{pies_f2041_nodate}.
These crystals belong to the monoclinic wolframite structure type (or triclinic one in the case of CuWO$_4$), with a distorted hexagonal close packing of the oxygen atoms and the metal atoms occupying one-fourth of the octahedral holes of the packing \cite{wells_structural_2012}.
They all are antiferromagnets, but the types of magnetic ordering are different. In the last decade, MnWO$_4$, and some related materials with partial substitution of Mn for Co, had attracted much interest in the multiferroic communty \cite{ptak_temperature-dependent_2012,meier_topology_2009,meier_second_2010,urcelay-olabarria_conical_2012,herrero-martin_direct_2015}. Several magnetic and structural phase transitions occur in MnWO$_4$ below \textit{T}$_N$ and the control of electrical polarization by the magnetic field was achieved \cite{taniguchi_ferroelectric_2006, hoffmann_time-resolved_2011}. This group of materials was also studied by using the dielectric spectroscopy methods \cite{bharati_electrical_1980,pullar_mgwo_2007}.
Many magnetic and nonmagnetic wolframites are actively studied due to their promising properties as cathode materials for asymmetric supercapacitors \cite{niu_simple_2013} and as hydrodesulfurization catalysts \cite{bi_niwo_2010}. Scintillating properties of wolframites also attract much attention \cite{mikhailik_scintillation_2007,annenkov_lead_2002}. Nevertheless, many other fundamental properties of \textit{A}WO$_4$ wolframites, e.g. such as the lattice and magnetic dynamics are still scarcely investigated.




In this paper, we report on the lattice and spin dynamics in single crystals of the nickel wolframite NiWO$_4$ studied by polarized Raman spectroscopy.
Recent studies of NiWO$_4$ powders using the synchrotron radiation in the infrared (IR)  \cite{kalinko_synchrotron-based_2016} and visible \cite{kuzmin_uv-vuv_2016} spectral range allowed to get some insights on polar IR-active phonons, as well as on electron excitation bands.
However, no magnetic excitations were found, probably due to the use of the powder samples, or to some other experimental limitations of the technique.
In contrast, we will show that we have succeeded to observe surprisingly rich low-energy spin-related excitations, regardless of the relatively simple magnetic structure of this material.
Perhaps single-crystal samples of good quality are crucial for experimental observation of spin dynamics in wolframites.
A series of magnetic excitation in the frequency range of 10-200 cm$^{-1}$ in both paramagnetic and antiferromagnetic (AFM) states were uncovered.
Our experimental observation are supported by the symmetry analysis, adopting a general approach, but our conclusions should be applicable to all magnetics with the same space group and magnetic propagation vector, e.g. CoWO$_4$.
The linear spin-wave theory (LSWT) calculations on the basis of observed and known from literature static and dynamic magnetic properties were used to derive spin-wave dispersion curves and to evaluate micromagnetic exchange and anisotropy parameters.



The paper is organized as follows.
Sec.~\ref{sec:exp} introduces samples growth and preparation procedure and experimental setups.
Magnetostatic measurements are introduced here as well.
In Sec.~\ref{sec:lattice_dynamics} the crystal structure and symmetry of the phonon modes along with experimental results of lattice dynamics are discussed.
Magnetic structure, magnetic symmetry and spin dynamics with interpretation and theoretical models are given in Section~\ref{sec:mag_struc_spin_dynamics}.
Section~\ref{sec:concl} summarized experimental and theoretical results along with conclusions.

\section{\label{sec:exp}Experimental details}

\subsection{\label{sec:exp_samples}Samples}
Single crystals of \Ni{} have been grown by the melt-solution growth technique, using sodium polytungstate as melt solvent, as described in \cite{schultze_zur_1967,oishi_growth_1996}.
Powders of \Ni{}, Na$_2$WO$_4$$\cdot$2H$_2$O and WO$_3$ were mixed in a ratio of 1~:~2~:~3, homogenized and heated to 1373~K for 24~h, cooled at a rate of 3~K/h to 1023~K and then cooled to room temperature within 12~h.
The crystals were separated from the flux with a solution of NH$_3$ in water.
The resulting crystals show well-developed morphology with  \{100\}, \{110\}, \{010\} and \{102\} forms.

All samples used for experiments were twin-free and carefully lapped to $\approx$3x2x0.4~mm plates to provide better cooling in the cryostat and polished to diminish the surface roughness scattering.

Due to the fact that the monoclinic angle $\beta$ is very close to 90$^{\circ}$, we used the pseudo-orthorhombic setting of the axes for Raman scattering polarization notation.
In magnetically ordered phase an alternative set of axes was added, \textit{Z} parallel to the antiferromagnetic vector \textbf{L} which lies in the \textit{ac}-plane at 19.4$^{\circ}$ in obtuse $\beta$ \cite{wilkinson_magnetic_1977}, \textit{Y}~$\parallel$~\textit{b} and \textit{X}~$\perp$~\textit{Z}.

\subsection{\label{sec:exp_raman}Raman spectroscopy}
Raman scattering spectra were measured in the range of 10--1600~cm$^{-1}$ with the use of a T64000 (Jobin-Yvon) spectrometer in the triple monochromator subtraction mode equipped with a liquid-nitrogen-cooled CCD camera.
All measurements were done in the back-scattering geometry for all required  polarization settings.
The line at 532~nm (2.33~eV) of a Nd:YAG-laser (Torus, Laser Quantum, Inc.)  with appropriate interference filter was used as the excitation source.
A 50x objective was employed both to focus the incident beam and to collect the scattered light.
Low-temperature spectra were recorded using a helium closed-cycle cryostat (Cryo Industries, Inc.) with temperature stability better than 1~K.
Single crystals of \Ni{} were mounted on the cold finger with the use of a silver paste.
Before the experiments the spectrometer was calibrated using the silicon Si-phonon line at 520.7 cm$^{-1}$.

\subsection{\label{sec:exp_magnetic}Magnetic measurements}
\begin{figure}
\includegraphics[width=85mm]{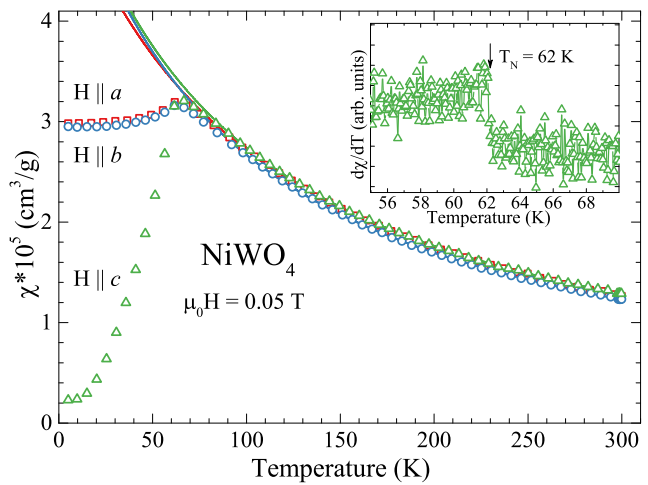}
\caption{\label{fig:suscep} 
(Color online) Magnetic susceptibility for the three main crystallographic axes. Only every 20th point is shown. Continuous lines depict the Curie-Weiss fit according to Eq.~(\ref{eq:Curie-Weiss}). Inset shows the first derivative of the H $\parallel c$ susceptibility which confirms the phase transition.} 
\end{figure}

Static magnetic measurements were done at the PPMS 14T system (Quantum Design).
M(T) and M(H) dependences up to 14~T were recorded along the three pseudo-orthorhombic crystallographic axes.
Figure~\ref{fig:suscep} shows the  magnetic susceptibility data measured along the three main crystallographic axes.
The drastic change of $d\chi/dT$ vs $T$ for H$\parallel$\textit{c} below $T_N$ = 62~K clearly indicates the long-range antiferromagnetic (AFM) transition.
The easy  axis of antiferromagnetic ordering is close to the \textit{c}-axis.

In the paramagnetic regime the magnetic susceptibility follows the Curie-Weiss law in the range of 95--300~K, and this range was used for fitting the experimental data by applying Eq.~(\ref{eq:Curie-Weiss}), taking into account a small temperature-independent term $\chi_{0}$ containing the diamagnetic and Van Vleck paramagnetic contributions:

\begin{equation}
\label{eq:Curie-Weiss}
\chi = C / (T + \Theta) + \chi_{0}
\end{equation}

Deviation of the susceptibility from the Curie-Weiss fit in the temperature range of 62--95~K is evidently caused by the short-range magnetic ordering.
Paramagnetic Curie temperatures were determined using the fitting procedure and the following values were found: $\Theta_a$ = -106~K,  $\Theta_b$ = -85~K and $\Theta_c$ = -91~K.
Negative values are characteristic marks of the dominant AFM interaction.
The Curie constants yield the magnetic moments in Bohr magnetons $\mu_{a,eff}$ = 3.19$\mu_{B}$, $\mu_{b,eff}$ = 2.93$\mu_{B}$, $\mu_{c,eff}$ = 3.01$\mu_{B}$.
These values are close to the theoretical spin-only value with \textit{S}=1, $\mu = 2.83$ and lie within the range of 2.9--3.3 for Ni$^{2+}(3d^8)$ ions in the octahedral coordination.
It is admissible to attribute higher values of magnetic moments to orbital contribution of excited states, e.g. \textit{t}$_{2g}^5$\textit{e}$_g^3$.
These effective moments allow us to calculate the \textit{g}-factors from the relation $\mu =g[S(S+1)]^{1/2}$, which are 2.25, 2.07 and 2.13 for the three crystallographic axes, respectively.
These numbers are reasonably close to the typical \textit{g}-factor values of the Ni$^{2+}$ ion \cite{hwang_successive_2012}.
However the determined \textit{g}-factors  differ from the values reported in Ref.~\cite{eremenko_low_1974}, where the authors claimed that the largest value of \textit{g} = 2.26 lies along the $z$-axis, which deviates from the $c$-axis by 15$^{\circ}$.
We note, that those estimates of the \textit{g}-factors were based on the behavior of magnetic excitations in an applied field.
For a known Curie-Weiss temperature the frustration index $f = |\Theta|/T_N \approx 1.5$ can be calculated, where the $\Theta$ is an averaged value for the all three axis.
This estimate points to a negligible role of the frustration for the magnetic structure of \Ni{}.
No ferromagnetic moment and no spin-orientation transition were observed in measurements in the field H up to 14~T along the three crystallographic axes for temperatures between 5 and 60~K. 
A spin-flop transition at H = 17.5~T was reported in Ref.~\cite{dudko_low_1980,eremenko_domain_1979}.

\section{\label{sec:lattice_dynamics}Lattice dynamics}
\subsection{\label{sec:cryst}Crystal structure and phonon modes}
\Ni{} belongs to the wolframite-type structure, space group $P2/c$, No.13, point group $2/m$. This structure is regarded as a deformed scheelite structure typical for such materials as CaWO$_4$, PbMoO$_4$, PbWO$_4$ \cite{pies_f2041_nodate}.
Crystal lattice parameters are listed in TABLE~\ref{tab:cryst} according to Ref. \cite{weitzel_kristallstrukturverfeinerung_1976}.
The monoclinic angle $\beta$ = $90.03^{\circ}$ is very close to $90^{\circ}$.

The symmetry analysis of the crystal unit cell of \Ni{} yields 36 phonon modes, see Eq.~(\ref{eq:modes}), namely 18 even Raman active ($8A_g+10B_g$) modes, 15 odd infrared ($7A_u+8B_u$) modes and three acoustic ones ($1A_u+2B_u$).
Here we will focus on the Raman modes and the corresponding Raman tensors are shown in Eq.~(\ref{eq:raman_tensors}).
It is clearly seen, that $A_g$ modes can be observed in all diagonal polarizations, e.g. $a(bb)\bar{a}$, $b(cc)\bar{b}$, etc., as well as in crossed off-diagonal ones $y(xz)\bar{y}$ and $y(zx)\bar{y}$.
The $B_g$ modes can be found only in crossed $c(ab)\bar{c}$ and $a(bc)\bar{a}$ polarizations. Obviously, even for the allowed polarizations some modes may not be seen, due to the weakness of the relevant elements in the Raman tensors.

\begin{table}
\caption{\label{tab:cryst}Crystal parameters of \Ni{} \cite{weitzel_kristallstrukturverfeinerung_1976}}.
\begin{ruledtabular}
\begin{tabular}{c|cccc}
Space group \#13&Atom and&\multicolumn{3}{c}{Reduced coordinates}\\
\textit{P2/c} (C$_{2h}^{4})$&Wyckoff position&x&y&z\\
\hline
a = 4.5992 \AA&Ni (2\textit{f})&0.5000&0.6616&0.2500\\
b = 5.0606 \AA&W (2\textit{e})&0.0000&0.1786&0.2500\\
c = 4.9068 \AA&O1 (4\textit{g})&0.2241&0.1105&0.9204\\
$\beta= 90.03^{\circ}$&O2 (4\textit{g})&0.2644&0.3772&0.3953
\end{tabular}
\end{ruledtabular}
\end{table}

\begin{equation}
\label{eq:modes}
\Gamma = 8A_g + 10B_g + 8A_u + 10B_u
\end{equation}

\begin{equation}
\label{eq:raman_tensors}
A_g =
\begin{pmatrix}
  a & 0 & d \\
  0 & b & 0 \\
  d & 0 & c
\end{pmatrix},
B_g = 
\begin{pmatrix}
  0 & e & 0 \\
  e & 0 & f \\
  0 & f & 0
\end{pmatrix}
\end{equation}

\subsection{\label{sec:results_raman}Lattice scattering}
\begin{figure}
\includegraphics[width=85mm]{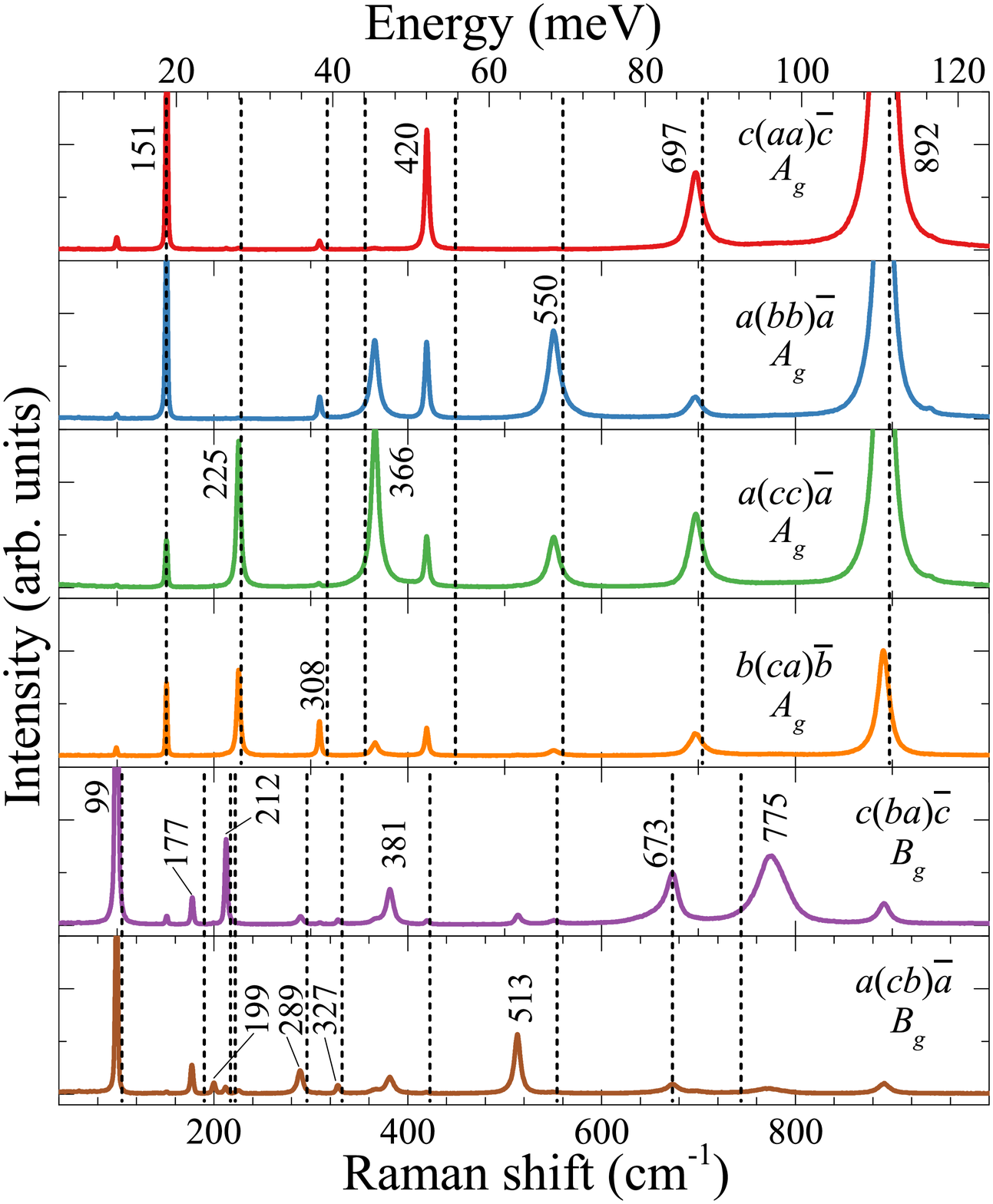}
\caption{\label{fig:room}
(Color online) Room temperature Raman scattering spectra of \Ni{} for parallel and crossed polarizations.
Vertical dashed lines show \textit{ab initio} calculated frequencies, according to Ref.~\cite{kuzmin_first-principles_2011}.
Numbers represent experimentally registered phonon frequencies.}
\end{figure}

As discussed above 18 Raman-active phonon modes are expected in \Ni{}, namely 8$A_g$ and 10 $B_g$ modes.
For distinguishing modes of different symmetries, measurements were done using the complete set of incident-outgoing polarizations and all expected phonons were observed.
Fig.~\ref{fig:room} shows room temperature phonon spectra.
Depolarization in optical elements and small misalignment of light polarization and crystallographic axes lead to small leak of forbidden modes in some spectra.
Non-equivalency in the Raman-tensors elements, see Eq.~\eqref{eq:raman_tensors} leads to different intensities for the modes of the same type, which often happens in low-symmetry crystals.
For example, the $A_g$ phonon at 225 cm$^{-1}$ is clearly visible only in the (\textit{cc}) and (\textit{ca}) polarizations.
The phonon frequencies, intensities and full-width at half maxima (FWHM) were extracted by fitting the Raman spectra to Voigt profiles \cite{wojdyr_fityk:_2010}.
Frequencies of all 18 observed modes are listed in Table~\ref{tab:phon}, along with the results of hybrid functionals DFT calculations \cite{kuzmin_first-principles_2011}.
FWHMs of the latter phonons are also presented in Table~\ref{tab:phon}. 
Assignments of the vibrational modes can be adopted from related research for Fe and Co doped MnWO$_4$ \cite{maczka_lattice_2011}, since both crystals are isostructural.

\begin{table}
\caption{\label{tab:phon}
Phonon modes, their polarizations, experimental phonon frequencies (cm$^{-1}$) and their FWHM (cm$^{-1}$).
The fourh column shows the DFT results from Ref.~\cite{kuzmin_first-principles_2011}.
}
\begin{ruledtabular}
\begin{tabular}{c|c|c|c|c}
Mode&Polarization&Frequency&DFT\cite{kuzmin_first-principles_2011}&FWHM\\
\hline
$B_g$&(\textit{ba})&99&105&2.7\\
$A_g$&(\textit{aa}),(\textit{bb})&151&151&2.6\\
$B_g$&(\textit{cb}),(\textit{ba})&177&190&2.4\\
$B_g$&(\textit{cb})&199&217&3.7\\
$B_g$&(\textit{ba})&212&222&3.2\\
$A_g$&(\textit{cc})&225&228&4.7\\
$B_g$&(\textit{cb})&289&296&6.7\\
$A_g$&(\textit{ca})&308&317&3.0\\
$B_g$&(\textit{cb})&327&332&4.1\\
$A_g$&(\textit{cc})&366&356&10.1\\
$B_g$&(\textit{ba})&381&423&8.1\\
$A_g$&(\textit{aa})&420&449&4.3\\
$B_g$&(\textit{cb})&513&554&7.5\\
$A_g$&(\textit{bb})&550&560&13.4\\
$B_g$&(\textit{ba})&673&673&21.6\\
$A_g$&(\textit{aa}),(\textit{cc})&697&704&15.7\\
$B_g$&(\textit{ba})&775&744&36.0\\
$A_g$&(\textit{aa})&892&897&14.3\\
\end{tabular}
\end{ruledtabular}
\end{table}

All Raman scattering experiments for the (\textit{bb}), (\textit{cb}), (\textit{ac}), and (\textit{cc}) polarizations were done in the temperature range of 10--300~K, and for the (\textit{aa}) and (\textit{ba}) polarizations in the range of 10--220~K. 
Temperature measurements were done in the region of 12-620~cm$^{-1}$, where only 14 of the total 18 phonon modes are observable.
We note that no emergence of new phonons was observed down to 10~K which proves the absence of any structural phase transition.
However, noticeable hardening of the most phonon lines are observed at temperatures below $T_N$ thus confirming non-vanishing coupling between lattice and magnetic subsystems through spin-phonon interaction.

\section{\label{sec:mag_struc_spin_dynamics}Magnetic structure and spin dynamics}
\subsection{\label{sec:sub_mag_struc}Magnetic structure}
In this subsection the magnetic structure of \Ni{} will be discussed because its reliable identification is essential for the analysis of physical properties.
For doing this, we will apply two techniques known as i) the magnetic space group analysis, and ii) the irreducible representation analysis.

\subsubsection{\label{sec:MSG}Magnetic space group analysis}
It is known from earlier studies \cite{wilkinson_magnetic_1977} that magnetic structure of \Ni{}, consisting of Ni$^{2+}$ \textit{S}=1 ions, can be described with a single magnetic propagation vector \textbf{k}=(1/2,0,0).
This information along with the known space group and crystal parameters of the paramagnetic phase is sufficient for the symmetry analysis within the magnetic space group (MSG) approach according to the MAXMAGN code at the Bilbao crystallographic center \cite{perez-mato_symmetry-based_2015}.
There is no reason to assume that the magnetic group is lower then the k-maximal one, so we will begin the analysis with these assumptions.

\begin{figure}
\includegraphics[width=85mm]{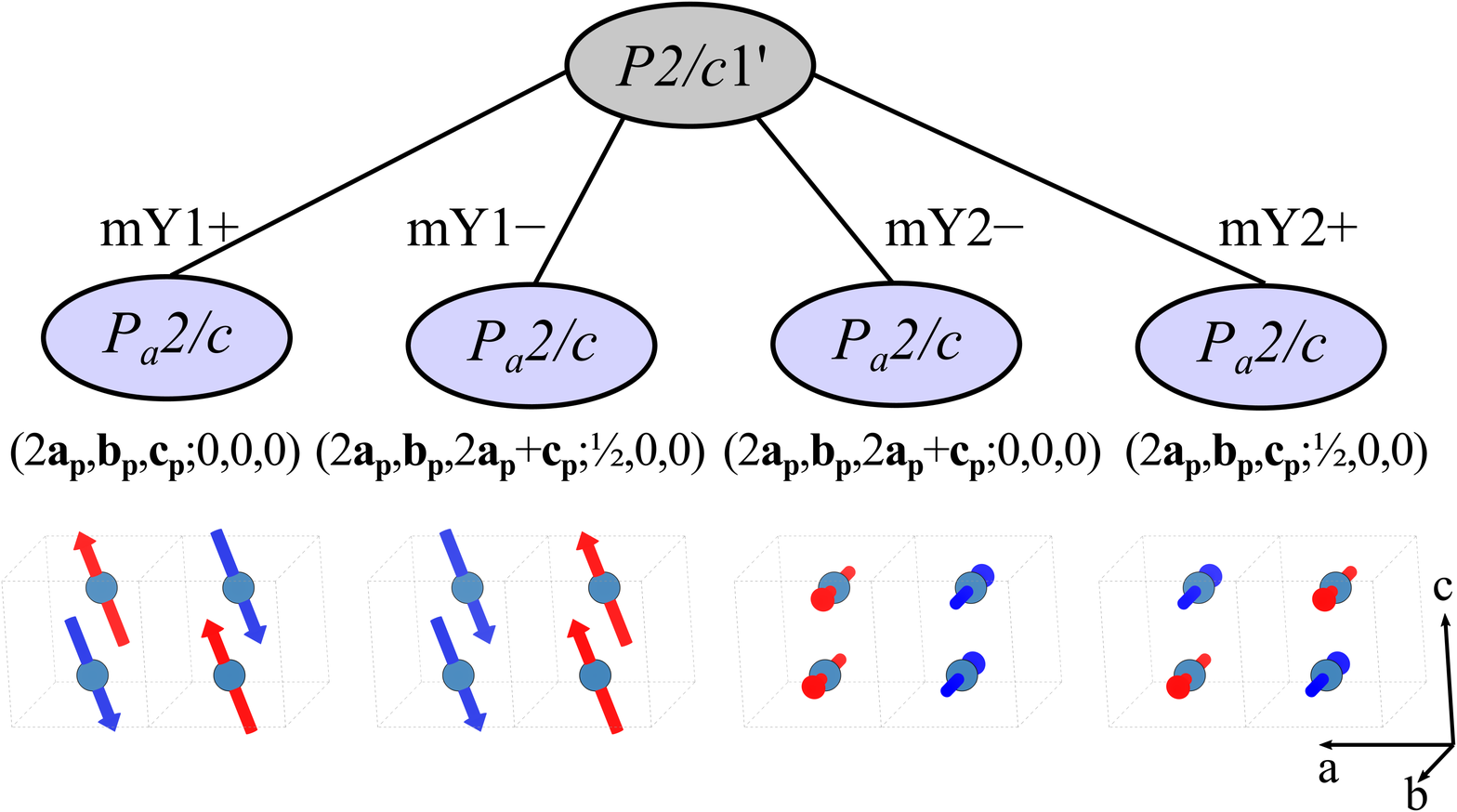}
\caption{\label{fig:sub} 
(Color online) The four admissable k-maximal magnetic subgroups with the propagation vectors \textbf{k}=(1/2,0,0) for the 2\textit{f} Ni sites for the gray paramagnetic parent group with the corresponding \textit{irreps}.
Transformation matrices to standard setting are given in parenthesis.
Index (G:Gmag) of the subgroups is two.
Magnetic unit cells are given in parent-like setting and are doubled along the \textit{a}-axis with respect to the crystallographic unit cell. At the two first images from the left the spins lie in the (\textit{ac})-plane, at the two next ones along the \textit{b}-axis.}
\end{figure}

The k-maximal magnetic subgroups for the chosen parent gray group are shown in FIG.~\ref{fig:sub}.
Images are visualized with the use of the VESTA program \cite{momma_vesta_2011}.
Only four magnetic k-maximal groups exist, as shown, and only one of them is relevant to \Ni{}.
Two of them allow only the $M_y$ component of the magnetization, while two others allow simultaneous presence of the $M_x$ and $M_z$ components.
From neutron-diffraction experiments it is known that magnetic moments of the Ni ions lie within the (\textit{ac})-planes, and are coupled ferromagnetically along the [001] direction.
Thus the second from the right magnetic space group $P_a2/c$ (No.13.70 in BNS notation) and $P_{2a}2/c$ (No.13.6.82 in OG notation) describes the magnetic symmetry of the aniferromagnetic phase of \Ni{}.
This magnetic space group constrains moments within the (\textit{ac})-plane, and spins have only two degrees of freedom.
We note that no splitting of atomic sites due to the magnetic ordering is predicted, and no electrical polarization are expected as well.
The index of the MSG with respect to the parent group is two, hence only trivial AFM domains with inversed orientation of spins are expected. 

\subsubsection{\label{sec:represent}Representation analysis}
The representation analysis was applied to \Ni{} with the use of the ISODISTORT code \cite{campbell_isodisplace:_2006}.
The magnetic propagation vector \textbf{k}=(1/2,0,0) corresponds to the Y-point in the Brillouin zone.
Accordingly, the four \textit{irreps} were found in total: i) mY2+ and mY2- with the only one degree of freedom for spins along the \textit{b}-axis; ii) mY1+ and mY1- with the two degrees of freedom with spins restricted within the (\textit{ac})-plane.
These irreducible representations are depicted in FIG.~\ref{fig:sub}.
In the case of the one-dimensional (1D) representations, results are equivalent to the MSG analysis.
Thus, the determined magnetic symmetry of \Ni{} corresponds to the $P_a2/c$ magnetic space group which is equivalent with the mY1-\textit{irrep}.

\subsubsection{\label{sec:exchange}Exchange structure}
\begin{figure}
\includegraphics[width=85mm]{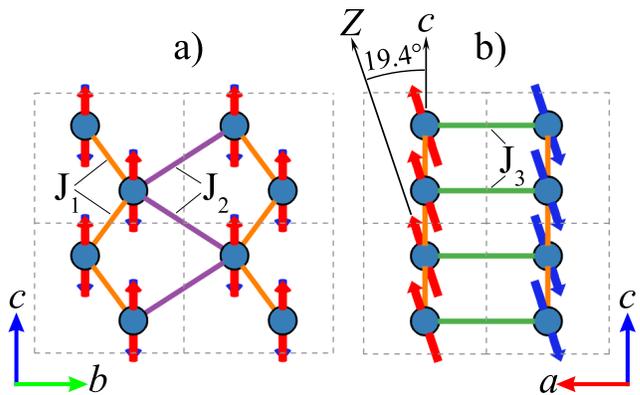}
\caption{\label{fig:exch}
(Color online) Magnetic structure of \Ni{} compatible with experiment and symmetry analysis given in Sec.~\ref{sec:sub_mag_struc}.
Exchange-coupling paths used in calculations are shown by solid lines.
Note, that eight crystal unit cells are shown to depict all exchange paths.}
\end{figure}

Starting from already known collinear antiferromagnetiс structure \cite{wilkinson_magnetic_1977}, in which the AFM vector lies in the \textit{ac}-plane and deviates from the \textit{c}-axis by 19.4$^{\circ}$, it is possible to analyze exchange paths.
Only a minimum number of magnetic couplings required to construct the 3D magnetic structure was taken into account.
The corresponding three interactions are shown in FIG.~\ref{fig:exch}.
Ferromagnetic ordering within the crystallographic unit cell and its doubling along the \textit{a}-axis suggest the sign of the following exchange constants, namely, ferromagnetic J1, J2, and antiferromagnetic J3, which finally create the antiferromagnetic structure.

The super-exchange path J1 involves [NiO$_6$] octahedra edges connected via O2 ions, with Ni-O-Ni distances 2.06144 and 2.06759~\AA~  and the angle of $\approx$~95.7$^{\circ}$ which approaches 90$^{\circ}$.
The sign of the coupling constant J1 is in agreement with the Goodenough-Kanamori-Anderson rules \cite{goodenough_magnetism_1966}, which predict the FM interaction for this geometry.
The J1 constant couples magnetic Ni ions in zig-zag-like FM chains running along the \textit{c}-axis.
These chains, coupled by the FM super-super exchange J2 (Ni--Ni distance is 4.549 \AA), form 2D-like FM sheets, which, in their turn, are coupled antiferromagnetically through J3 (Ni--Ni distance is 4.599 \AA) forming the 3D AFM structure of \Ni{}.

\subsection{\label{sec:raman_magnetic}Magnetic scattering}
\subsubsection{\label{sec:sub_raman_magnetic}Magnetic scattering}
\begin{figure*}
\includegraphics[width=178mm]{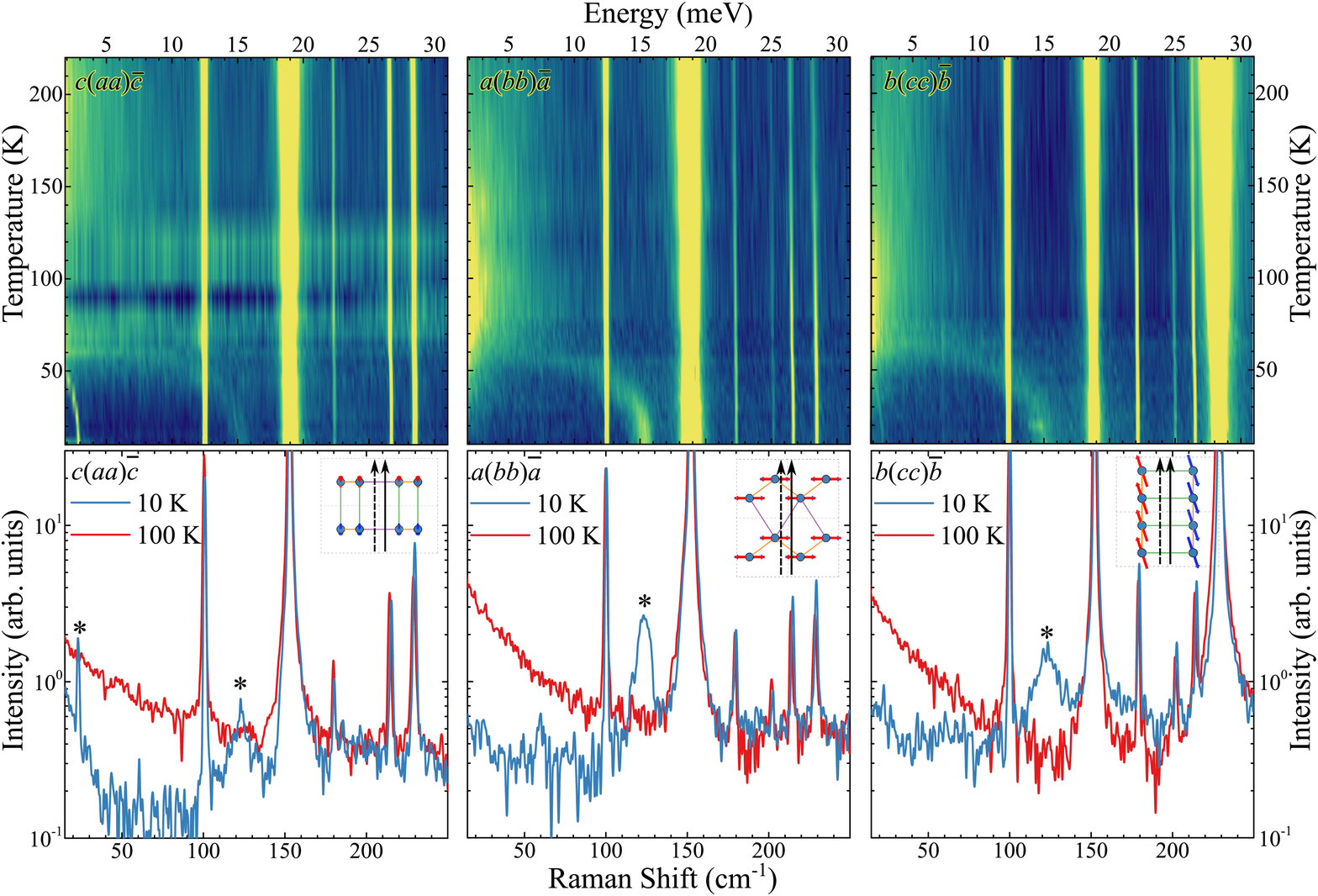}
\caption{\label{fig:temp_mag_parallel}
(Color online) 
Upper plots show temperature maps of the Raman scattering spectra for the three diagonal (\textit{aa}), (\textit{bb}), and (\textit{cc}) polarizations in the range of 10--220~K. Lower plots show the relevant spectra in the logarithmic scale above and below $T_N$=62~K at \textit{T} = 10 and 100~K.
The polarization vectors of incoming (dashed) and outgoing (solid) photons with respect to magnetic structure are shown as insets.
Magnetic excitations are marked with asterisks.}
\end{figure*}

\begin{figure*}
\includegraphics[width=178mm]{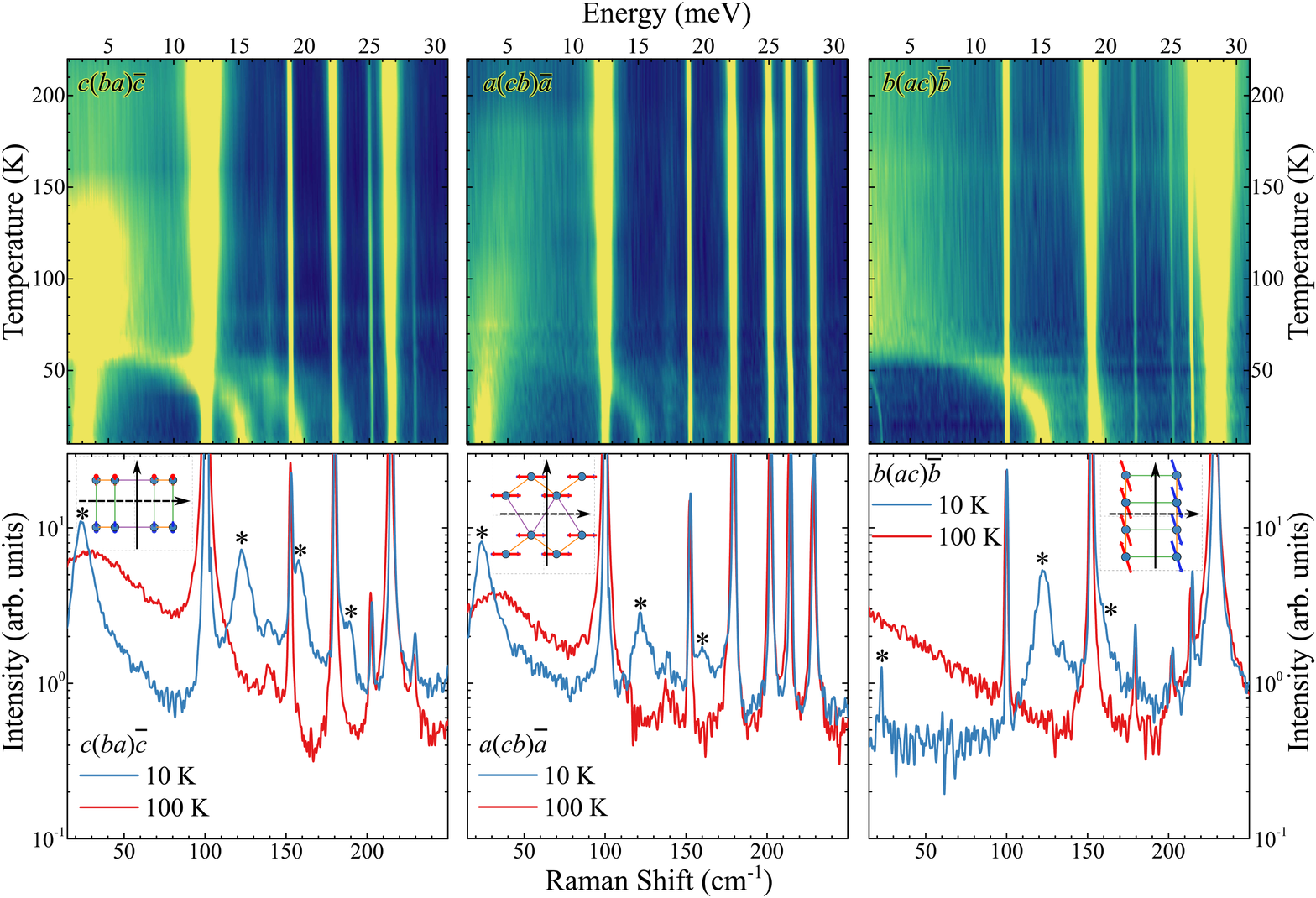}
\caption{\label{fig:temp_mag_crossed}
(Color online) Upper plots show maps of the Raman scattering spectra in off-diagonal crossed polarizations in the temperature range of 10--220~K.
Lower plots show the Raman spectra in the logarithmic scale above and below $T_N$=62~K at temperatures 10 and 100~K.
The polarization vectors of incoming (dashed) and outgoing (solid) photons with respect to the magnetic structure are shown as insets.
Magnetic excitations are marked with asterisks.
}
\end{figure*}

Hereafter the observation of magnetic Raman scattering will be discussed. Typically, the energy scale of magnetic scattering is on the order of the transition temperature from paramagnetic to antiferromagnetic state since it is roughly proportional to the exchange couplings in the system.
We remind that in \Ni{} $T_N$=62~K~$\approx$~45~cm$^{-1}$.
Several spectral features unambiguously related to magnetic excitations were observed over a much higher frequency range up to 200~cm$^{-1}$, and within the temperature range from 10 to 220~K much higher $T_N$. 
Figures~\ref{fig:temp_mag_parallel} and ~\ref{fig:temp_mag_crossed} summarize the experimental results obtained in different polarizations below and above $T_N$.
No structural transition is known in \Ni{}.
Moreover, the lowest electronic states lie above 1~eV with the band gap in nickel oxides above 4~eV \cite{pisarev_lattice_2016}.
These properties allow us to exclude structural and electronic mechanisms from discussion and assign the observed features showing strong correlations to the N\'eel temperature only to magnetic excitations. 

The first noticeable feature is an intense quasi-elastic scattering observed in several polarizations (e.g. see (\textit{bb}) polarization in Fig.~\ref{fig:temp_mag_parallel}) as a strong shoulder of the excitation line in the paramagnetic region above \textit{T}$_N$.
The shape of this quasi-elastic tail is well described by the Lorentzian profile, suggesting major contribution of the fluctuations of the magnetic energy density \cite{yamada_light_1994}.
This quasi-elastic magnetic scattering is strongly suppressed when the long-range AFM order is established.  
Besides that, at least four other different types of magnetic excitations are observed in \Ni{} which are distinguished on the basis of their different temperature behavior and polarization properties.

First of all, there is a weak very narrow ($\approx$1~cm$^{-1}$), in comparison to phonons excitations, magnetic mode at 22.5~cm$^{-1}$ which was observed only in the (\textit{aa}), (\textit{ac}) polarizations and as a very weak line in the (\textit{cc}) polarization.
It is quite probable that this mode might be active in other polarizations, e.g. (\textit{ba}),(\textit{cb})  but it is overshadowed by other more pronounced excitations.
This mode at 22.5~cm$^{-1}$ can be assigned to one of the lowest branches of the complex spin-wave spectra (acoustic magnon). First of all, this assignment is supported by its frequency, an extremely narrow line width, its temperature behavior which follows the Brillouin function, see Fig.~\ref{fig:mag_freq}, and, finally, by theoretical calculations presented in Sec.~\ref{sec:results_magnetic_structure}.
The symmetry of this mode, predominately observed in polarizations parallel to the main AFM exchange path, also suggests that it is related to the AFM resonance.
We add that two one-magnon modes with close frequencies were observed in the far-infrared studies of \Ni{} \cite{eremenko_low_1974}. 
However in our Raman studies only the higher-energy mode was observed either because of the selection rules or as a cause of too low intensity of another mode.
Nonzero energy of the acoustic spin-wave branch (AFM resonance) and additional zero-field splitting can be explained by the simultaneous presence of the easy axis and easy plane single-ion anisotropies.

Another strongly pronounced but very broad magnetic excitation was found at the frequency of 24 cm$^{-1}$ in crossed (\textit{ba}) and (\textit{cb}) polarizations in the close vicinity to previously mentioned one-magnon line.
To the best of our knowledge no report on existence of such a mode in \Ni{} is available in literature.
Though its frequency is very close, within 2 cm$^{-1}$, to that of the discussed above one-magnon mode, it demonstrates drastically different behavior.
First of all, its integral intensity is about two orders of magnitude larger and the FWHM is much broader.
This excitation does not soften, but even slightly hardens by a few cm$^{-1}$ when approaching $T_N$ from below (Fig.~\ref{fig:mag_freq}).
It survives up to 250~K where it overlaps with the paramagnetic quasi-elastic scattering, e.g. in the (\textit{ba}) polarization, resulting in a very broad and intense band.

The following tentative assumptions concerning the nature of this mode may be suggested: i) a two-magnon band, but it seems not to be very likely because the relevant process which simultaneously creates two magnons with \textbf{k} and \textbf{-k} and the resulting excitation should cover the range of at least twice the lowest one-magnon energy ($>2*17.9$~cm$^{-1}$). Moreover, the temperature behavior of this mode also excludes such hypothesis;
ii) Since the strong spin-phonon coupling takes place in \Ni{} we may assume that there is a coupling of the one of the AFM resonance modes with phonons thus resulting in extremely strong overdamping.
However this hypothesis does not explain temperature dependency of its frequency;
iii) Spin-gap mode due to the non-zero dimerization of Ni ions.
However, the dimerization typically favors ions with spins \textit{S}=1/2 coupled by a strong exchange interaction.
For example, a model of dimerized magnetic structure with a long range AFM order was suggested for a triclinic crystal CuWO$_4$ \cite{lake_dimer_1997} which is structurally close to \Ni{}.
A spin-dimer model was recently suggested for the Ni$^{2+}$ system Ni$_2$NbBO$_6$ \cite{narsinga_rao_antiferromagnetism_2015}.
iv) Haldane gap since \Ni{} can be described as a magnetic system of strongly coupled AFM chains of \textit{S}=1 spins.
Since first three assumptions have strong counterarguments, the latter one seems plausible.



The third observed excitation with the maximum at 123.3~cm$^{-1}$ (at \textit{T} = 10 K) is observed in all polarizations though in the (\textit{aa}) polarization it is very weak. Taking into account its frequency, FWHM and temperature behavior, we could tentatively assign this mode to a two-magnon excitation.
However, earlier studies showed that this mode is split in an applied magnetic field and have the magnetic-dipole character \cite{eremenko_low_1974}.
Such behavior cannot be attributed to the usual $S^z = 0$ two-magnon excitations.
Nevertheless, following the analysis given in \cite{fleury_scattering_1968}, this mode can be assigned to $S^z = 2$ transitions when a pair of magnons with opposite \textbf{k}-vectors propagates on one or another magnetic sublattice.
This type of second-order excitations, indeed, should be split in an applied magnetic field.
Another, more reliable assignment is the high energy branch of spin-wave spectra (optical magnon), see FIG.~\ref{fig:SW_disp}.
Spin-wave calculations in applied field shows (see Fig.~\ref{fig:SW_field}) that this modes is split twice as much as the lowest ones in agreement with \cite{eremenko_low_1974}.

An additional peak near 139.6~cm$^{-1}$ was found in the crossed \textit{ba}) and (\textit{cb} polarizations. It did not show any noticable temperature dependence and vanished in the background around 250~K.
Such a behavior, along with its extremely weak intensity suggests a phonon impurity mode, rather than magnetic-related excitation.

Two weak and broad bands at 157.9 and 189.2 cm$^{-1}$ were observed in \Ni{} which were not reported before in literature. 
These bands show weak and almost linear softening as shown in Fig.~\ref{fig:mag_freq}, and exist only at low temperatures $T < T_N$.
We suggest that they are related to intra-ionic transitions between the spin states of the Ni$^{2+}$ ions split by the exchange field.
We also do not exclude the possibility that these modes are related to the magnetic bound states \cite{lemmens_magnetic_2003}.

\begin{figure}
\includegraphics[width=85mm]{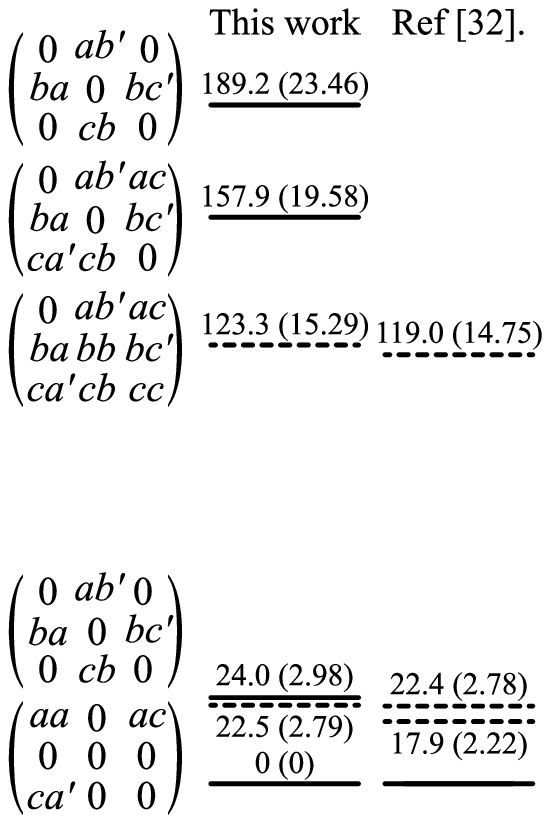}
\caption{\label{fig:levels}
(Color online) Scattering matrices and energy of the observed magnetic excitations in Raman scattering, in comparison with IR Ref.~\cite{eremenko_low_1974}.
Suggested spin-wave modes marked as dashed lines.
Numbers show excitations frequencies in cm$^{-1}$~(meV).
}
\end{figure}

Polarization properties of all observed magnetic excitations with reasonable intensity (more than 1 a.u. in Fig.~\ref{fig:temp_mag_parallel}) are summarized in Fig.~\ref{fig:levels} in form of scattering matrices.
It should be noted that primed off-diagonal elements were obtained suggesting symmetric form of these tensors.


Another striking feature is manifestation of interaction between different magnetic excitations which can be seen in temperature maps in Fig.~\ref{fig:temp_mag_crossed}.
Well pronounced individual modes are observed at $T=10$~K in the (\textit{cb}) polarization at 24~cm$^{-1}$ , and in the (\textit{ba}) polarization at 123.3~cm$^{-1}$.
But in the particular (\textit{ba}) polarization where both of them are present and overlap the strong coupling is observed in the vicinity of $T_N$.
A broad intense band can be seen up to 3$T_N$, and we may suppose that the quasi-elastic scattering interfere with this process as well.

There are several well pronounced signs of magnon-phonon interaction when the temperature-dependent magnetic mode at 123.3~cm$^{-1}$ ($T=10$~K) overlaps with the lowest-frequency phonon mode at 99~cm$^{-1}$ in the (\textit{ba}) polarization, where this mode has highest intensity.
This interaction is clearly revealed in a form of an asymmetric Fano-type resonance seen in Fig.~\ref{fig:temp_mag_crossed}.
If the 123.3~cm$^{-1}$ mode is related to the one- or two-magnon excitations having the exchange-scattering nature \cite{fleury_scattering_1968}, then it is reasonable to assume that the mechanisms of magnon-phonon coupling are due to the modulation of exchange interaction by this particular phonon mode. Previously it was assigned to a translational motion of tungsten ions \cite{ptak_temperature-dependent_2012}.
Such interaction of spin-waves with phonons, as well as already mentioned spin-phonon interaction undoubtedly requires further studies.


\subsubsection{\label{sec:results_magnetic_structure}Calculations of the spin-wave spectra}
Results of our both static and dynamic magnetic experiments allow us to construct exchange model and evaluate micromagnetic parameters such as particular exchange and single-ion anisotropy constants.
Calculations of the spin-wave spectra within the linear spin-wave theory (LSWT) with admitting the relaxation of the magnetic structure were done with the use of the SpinW library \cite{toth_linear_2015}.

\begin{figure}
\includegraphics[width=85mm]{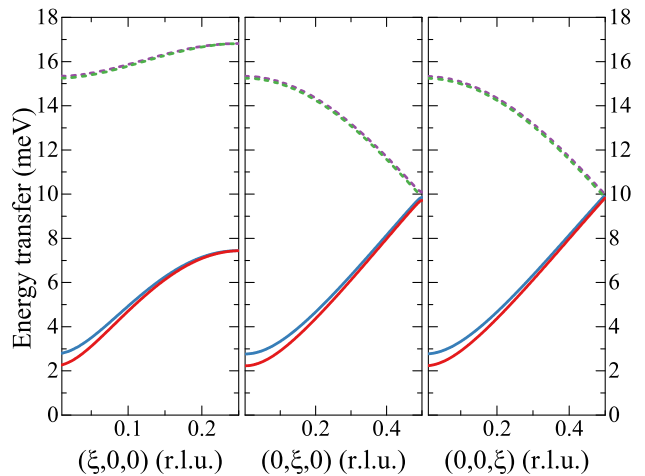}
\caption{\label{fig:SW_disp}
(Color online) Calculated spin-wave dispersions along the main crystallographic axes.
Parameters used for calculation are listed in Table~\ref{tab:micromag_parameters}.
Dashed lines represent the high-energy branch.
}
\end{figure}

To determine micromagnetic parameters we take into account the following experimental observations: i) Magnetic structure with deviation of magnetic moments from the \textit{c}-axis by 19.4$^{\circ}$ according to \cite{wilkinson_magnetic_1977}.
ii) Frequencies of one-magnon (AFM resonance) modes (17.9 and 22.4~cm$^{-1}$) according to \cite{eremenko_low_1974}.
iii) Additional frequency shift of these modes by -1 and +1~cm$^{-1}$ in the applied magnetic field of 1.8~T along the \textit{Z}-axis for the low and high frequency modes, respectively.
iv) The spin-flop transition at 17.5 T \cite{eremenko_domain_1979}.
v) Frequency of 123.3 cm$^{-1}$ mode, suggesting that it corresponds to the high energy zone-center mode, see Fig.~\ref{fig:SW_disp}.
vi) Magnon energy at the Brillouin-zone boundary equals to 60~cm$^{-1}$ as measured by the optical sideband absorption \cite{skorobogatova_effect_1972}.

Magnetic moments of the Ni ions are restricted to the \textit{ac}-plane as was previously mentioned, thus this plane is treated as an easy-plane.
Direction of the easy-axis anisotropy within this plane was chosen along the AFM vector \cite{wilkinson_magnetic_1977}, which is close to the longest Ni-O bonds within the [NiO$_6$] octahedra \cite{weitzel_kristallstrukturverfeinerung_1976}.
It should be noted, that the energy of the zone-center modes is much lower, than that in the copper wolframite CoWO$_4$ \cite{gredescul_magnetization_1972,eremenko_low_1974}, thus indicating a smaller single-ion anisotropy in \Ni{} (taking into account relatively close values of $T_N$ for both crystals).


Additional frequency shift of magnon excitations and presence of the spin-flop transition were also simulated by applying the magnetic field along the \textit{Z}-axis.
The micromagnetic parameters were determined by minimizing the squared differences of observed and calculated spin-wave excitations.
The particle swarm optimization (PSO) algorithm was used to avoid local minima.
Comparison of the calculated values with experimental ones along with evaluated parameters are summarized in Table~\ref{tab:micromag_parameters}.
\begin{table}
\caption{\label{tab:micromag_parameters} Experimentally observed and calculated spin-wave modes frequencies. Evaluated exchange and single-ion anisotropy constants (in meV).
Note that J $<$ 0 and J $>$ 0 stand for the FM and AFM coupling, respectively.}
\begin{ruledtabular}
\begin{tabular}{c|c|c|c|c}
Observables & \makecell{Exp. \\ values} & Calculated & \makecell{Evaluated \\ parameters}   & \makecell{Model \\ fit}\\
\hline
\makecell{Angle between \\ Z and c\footnote{Ref.~\cite{wilkinson_magnetic_1977}}} & 19.4$^{\circ}$ & 19.4$^{\circ}$ & J1 & -1.166 \\
\makecell{Low AFMR \\ mode (in field)\footnote{Ref.~\cite{eremenko_low_1974}}} & \makecell{17.9 \\ (16.9)}   & \makecell{17.99 \\ (17.42)} & J2 & -1.168 \\
\makecell{High AFMR \\ mode (in field)\footnotemark[2]} & \makecell{22.4 \\ (23.4)} & \makecell{22.31 \\ (22.91)} & J3 & 3.501 \\
Upper branch & 123.3 & 123.32 & D$_{plane}$ & 0.096 \\
Boundary mode \footnote{Ref.~\cite{skorobogatova_effect_1972}} & 60 & 59.56 & D$_{axial}$ & -0.171 \\
\end{tabular}
\end{ruledtabular}
\end{table}

It should be noted that since the Raman scattering is a probe for excitations close to the center of the Brillouin zone (ignoring second- and higher-order processes), in most cases it is complicated or even impossible to obtain a unique set of exchange and anisotropy parameters.
For solving this problem usually an inelastic neutron scattering experiments for several directions in the Brillouin zone are required for verifying or refining the exchange parameters.


\begin{figure}
\includegraphics[width=85mm]{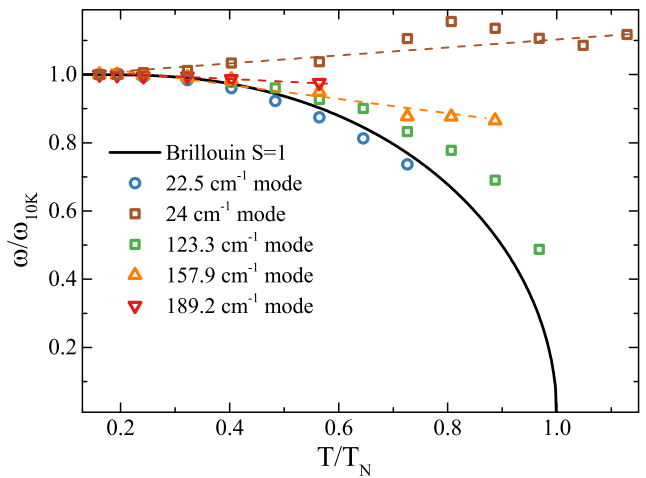}
\caption{\label{fig:mag_freq} (Color online) Normalized frequency dependencies vs normalized temperature of the observed magnetic excitations in comparison with the Brillouin function for \textit{S}=1 (solid line). Dashed lines are guides to the eye.}
\end{figure}

\begin{figure}
\includegraphics[width=85mm]{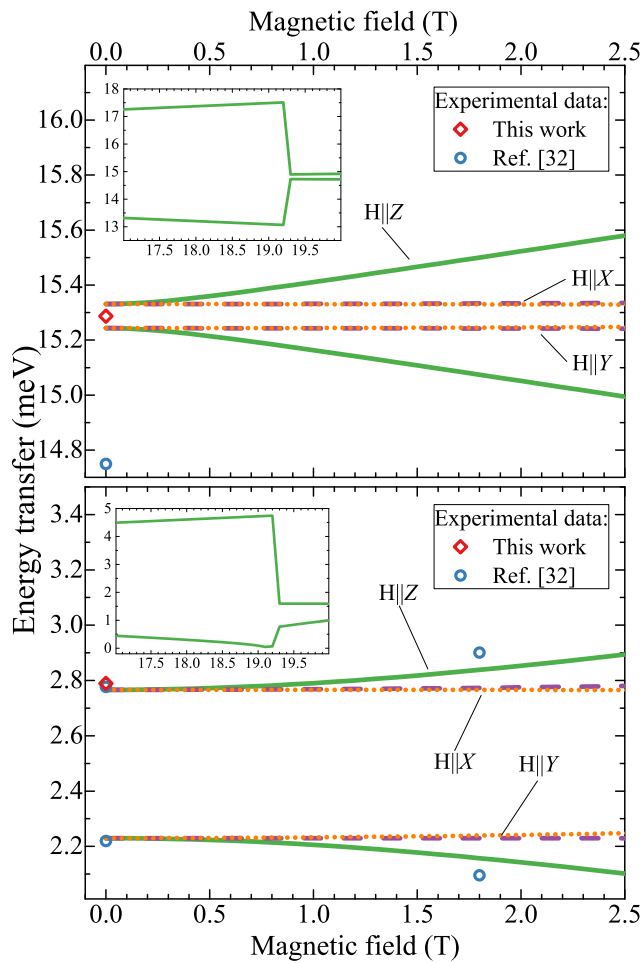}
\caption{\label{fig:SW_field}
(Color online) a) Calculated magnetic field dependence of the high-energy modes in the $\Gamma$ point for the field along \textit{X}, \textit{Y} and \textit{Z} axis.
b) Field dependence of the low-energy modes;
These results can be directly compared with those in Ref.\cite{eremenko_low_1974}.
Insets show spin-flop transition region for the both spin-wave branches.}
\end{figure}

Frequencies of observed magnetic excitations in \Ni{} and those reported in Ref.~\cite{eremenko_low_1974} are summarized in Fig.~\ref{fig:mag_freq}.
Following the results from Sec.~\ref{sec:sub_raman_magnetic} the 123.3~cm$^{-1}$ mode is assigned to high frequency zone-center excitations (optical magnon).
In the proposed model this mode is split even in zero applied magnetic field, however the splitting is less than 0.1~cm$^{-1}$ and cannot be resolved in our experiments.
Interesting to note, that since the magnon energy at the zone boundary is close to 60~cm$^{-1}$ one can expect appearance of a broad two-magnon band at about 120~cm$^{-1}$ whose broadening arises due to the magnon-magnon interaction.
In our experiments an extremely broad feature was observed in the (\textit{ba})-polarization as a background contribution in the range of 80--200~cm$^{-1}$.

Action of the magnetic field was simulated by applying it along $X$, $Y$ and $Z$ directions relaxing magnetic structure at each point and calculating spin-wave frequencies.
Results are shown in Fig.~\ref{fig:SW_field}, where shift of the spin-wave modes in the field up to 2~T can be directly compared to experimental data in Ref.~\cite{eremenko_low_1974}.
At higher fields the spin-flop transition can be seen as an abrupt change of magnon frequencies in both branches, which is shown as insets in Fig.~\ref{fig:SW_field}.
Calculated values of the critical field $H_{sf}$=19.2~T are in acceptable agreement with experimental data \cite{eremenko_domain_1979}.

We also should note that renormalization of the magnetic moments should be applied due to a rather small value of the quantum spin number \textit{S}=1, which can also results in modification of exchange parameters.
Thus we must conclude that the developed linear spin-wave model can explain only part of the experimentally registered magnetic excitations.
We suppose that further experiments, e.g. inelastic neutron scattering combined with more sophisticated theoretical models should be applied to \Ni{} for complete understanding of the excitations spectra and their temperature behavior.

\section{\label{sec:concl}Conclusions}
In conclusion, lattice and magnetic dynamics of the \Ni{} single crystals are studied with the use of polarized Raman scattering in the temperature range of 10--300~K was studied.
All possible phonon modes were observed and identified.
In the ordered phase and partly above critical temperature surprisingly rich magnetic excitation spectra were observed in the low-frequency region $<$~200~cm$^{-1}$.
Spin-wave model with three exchange couplings and combination of easy-plane and easy-axis anisotropy is shown to provide good description of magnon excitations. 
Additional previously unexplored magnetic excitations were observed and tentatively assigned to Haldane gap and single-ion excitations.
Obtained micromagnetic parameters provide the opportunity to further theoretical insights of the magnetic dynamics of wolframites and give a set of starting parameters to inelastic neutron scattering experiments for further investigations of spin dynamics.

\begin{acknowledgments}
Experimental and theoretical research was supported by the Russian Science Foundation grant No. 16-12-10456. The growth of single crystals was supported by the Deutsche Forschungsgemeinschaft (DFG) through SFB 1238.
\end{acknowledgments}

\bibliography{NiWO4}

\end{document}